\documentclass[reprint,amsmath,amssymb,prb]{revtex4-1}
\usepackage{amsmath}

\pdfoutput=1
\usepackage[pdftex]{graphicx}

\begin{document}
\preprint{}


\title{A new Wang-Landau approach to obtain phase diagrams for multicomponent alloys}

\author{Kazuhito Takeuchi}
\author{Ryohei Tanaka}
\author{Koretaka Yuge}
\affiliation{Department of Materials Science and Engineering, Kyoto University, Sakyo, Kyoto 606-8501, Japan}


\date{\today}

\begin{abstract}

We develop an approach to apply Wang-Landau algorithm to multicomponent alloys in semi-grand-canonical ensemble.
Although the Wang-Landau algorithm has great advantages over conventional sampling methods, there are few applications to alloys.
This is because calculating compositions in semi-grand-canonical ensemble using the Wang-Landau algorithm requires a multi-dimensional density of states in terms of total energy and compositions. However, constructing the multi-dimensional density of states is difficult.
In this study, we develop a simple approach to calculate the alloy phase diagram using Wang-Landau algorithm, and show that compositions in semi-grand-canonical ensemble require just some one-dimensional densities of states.
Finally, we applied the present method to Cu-Au and Pd-Rh alloys and confirmed that the present method successfully describes the phase diagram with high validity and accuracy.

\end{abstract}


\maketitle

\section{introduction}
\label{sec:introduction}
In alloy studies with first-principle calculation, estimating thermodynamic properties, especially free energy, is one of the great goals.
Thermodynamic integration (TI), which is based on Metropolis algorithm\cite{:/content/aip/journal/jcp/21/6/10.1063/1.1699114}, is one of the most widely used method to calculate thermodynamic properties and temperature-composition phase diagrams\cite{0965-0393-10-5-304, vandeWalle2002, VANDEWALLE2002539, vandeWalle2009266}.
In the simulations on a given lattice, cluster expansion\cite{SDG} (CE) formalism, which reconstructs the coarse-grained Hamiltonian from the outputs of density functional theory (DFT), plays an important role on alloys\cite{PhysRevB.88.094108, 0034-4885-71-4-046501, PhysRevB.71.054102} because CE describes the multibody interactions caused by the metallic bond.
Hence, the combination of DFT, CE and TI has been widely used for estimating alloy phase diagrams.

Although TI is a powerful method to calculate free energy, TI has a significant problem which suffers from phase transitions because of using Metropolis algorithm.
At 1st-order phase transitions, the tunneling barrier between coexisting phases increases exponetially by the Boltzmann factor.
This causes inaccurate estimation of phase transition points $(x, T)$ and also of the free energies near the transition points.
At 2nd-order phase transitions, due to critical slowing down, one could not estimate thermodynamic properties accurately near the transition points.
In alloy studies, these problems have been solved by looking for intersections of free energies between two metastable phases\cite{0965-0393-10-5-304, vandeWalle2002, VANDEWALLE2002539, vandeWalle2009266}.

Wang-Landau (WL) algorithm\cite{PhysRevLett.86.2050, PhysRevE.64.056101} is one of the most efficient and accurate method to obtain the density of states, which characterizes the thermodynamic properties of a considered system at equilibrium.
Let us explain the characteristics of the WL algorithm. Partition function, $Z$, is considered as:
\begin{equation}
	\label{eq:Z}
	Z = \sum_E W(E) \exp \left( - \frac{E}{k_{\rm B} T}  \right).
\end{equation}
Here, $W(E)$ is the density of states (DOS), $k_{\rm B}$ is the Boltzmann constant and $T$ is temperature.
Compared with the Metropolis algorithm\cite{:/content/aip/journal/jcp/21/6/10.1063/1.1699114}, the advantages of WL algorithm are
(i) to overcome the problems caused by 1st- and 2nd-order phase transitions,
(ii) to calculate Helmholtz free energy directly, and
(iii) once $W(E)$ is obtained, one could calculate free energy using Eq.~(\ref{eq:Z}) and $F=-k_{\rm B}T \ln{Z}$ at any temperatures.
These advantages are achieved by a random walker in the WL algorithm which covers the whole energy space and constructs the $W(E)$.

Although the WL algorithm has the great advantages over the conventional method, there are few applications to alloys.
This is because, in semi-grand-canonical ensemble for multicomponent alloys, a multi-dimensional DOS is typically required.
The WL studies on a multidimensional density of states\cite{Zhou_Wang_2006, Tsai_Uncovering_2008, Vogel_Generic_2013, Li_A_2014, Vogel_Exploring_2014, Vogel_Scalable_2014} shows some difficulties such as the connecting the pieces of $W(E)$ and computational costs.
Although the difficulties has been overcome by such as the multi-parallel framework\cite{Vogel_Generic_2013, Li_A_2014, Vogel_Exploring_2014, Vogel_Scalable_2014}, constructing a multi-dimensional density of states remains quite a difficult problem.

In this study, we suggest a method to construct the phase diagram for multicomponet alloys based on the WL algorithm, avoiding explicit construction of the multi-dimensional density of states.
We applied the present method to two binary alloys; Cu-Au and Pd-Rh that show the ordering and the phase-separation tendency respectively.
Through these two alloys, we confirmed that our method successfully calculate the phase diagram which captures the characteristic of Cu-Au and Pd-Rh,
and the combination of DFT, CE and WL is a prominent method to obtain alloy phase diagrams.

\section{methodology}
\label{sec:methodology}

First, in Sec.~\ref{sec:cluster_expansion}, we give a brief explanation of CE which describes total energy of alloy by the coarse-grained Hamiltonian.
Second, in Sec.~\ref{sec:wl_ce}, the conventional WL algorithm in canonical ensemble for alloys is presented.
Finally, in Sec.~\ref{sec:wl_sgc}, we show how to apply the conventional WL algorithm to multicomponent alloys, and why the conventional one is not suitable for multicomponent alloys.
As above, we present a new method based on the WL algorithm to handle multicomponent alloys.

\subsection{Cluster expansion}
\label{sec:cluster_expansion}
In CE, atomic configuration on a given lattice, $\sigma$, is described by a complete and orthogonal set of discrete basis functions.
Suppose that the occupation of element on lattice site $i$ is specified by Ising-like spin variable $S_i$.
CE introduces the cluster on lattice, $k$, e.g., points, pairs and triplets.
Especially in binary alloy, if the basis functions on a lattice point, $i$, are $\{ 1, S_i \}$ where $S_i$ has +1 or −1, so-called ``correlation function", $\xi_k$, is defined as the average of products of spin variables on $k$ over all symmetrically equivalent $k$ in $\sigma$.
Thus, configurational property, e.g., $E$, is completely represented via correlation functions and their coefficients:
\begin{equation}
\label{eq:CE}
	E_{\rm CE}(\sigma) = \sum_k V_k \xi_k (\sigma),
\end{equation}
where $V_k$ is called effective cluster interaction (ECI), which can be practically obtained by fitting the DFT formation energy.
Since in Eq.~(\ref{eq:CE}), $V_k$s are constant and only $\xi_k$s are variable, the formation energy for any configurations are easily obtained compared with DFT.
Thus, this coarse-grained Hamiltonian $E_{\rm CE}$ enables us to calculate a large number of energies that is required for the MC simulations.

\subsection{Wang-Landau algorithm in canonical ensemble}
\label{sec:wl_ce}
Suppose a random walker in configuration space.
If we consider the Ising spin model,
the moves of walker is often defined as single-spin flip, which may change its orientation.
However, if we consider the alloy system in canonical ensemble, single-spin flip method could not be used, because chemical potentials for each elements should be considered (unlike in Ising-spin model without a field) and therefore the compositions should be remain constant.
Thus, in order to fix the compositions, a pair of spins exchange, where a pair of spins attempt to exchange positions, is often used in alloy system.
Through the repeating update on the configuration, we obtain the time series of the random walker which corresponds to the samples of configuration.
When the acceptance ratio is proportional to $\exp \left( - \frac{E}{k_{\rm B}T} \right)$, the time series corresponds to the samplings from the canonical distribution.

In the WL algorithm, the acceptance ratio is proportional to $1/W(E)$ enabling the random walker to move randomly in energy space.
After the acceptance or rejection trial, $W(E)$ is updated as $W(E) \to W(E) \times f$ where $f$ is a modification factor that is initially greater than $f_0 = e^1$.
At the same time, the histogram $H(E)$ is also incremented as $H(E) \to H(E) + 1 $.
When $H(E)$ becomes sufficiently ``flat", $f$ is reduced such as $f_{i+1} = \sqrt{f_i}$ and all histogram bins are reset to zero.
Although there are many definitions of flatness of $H(E)$, in this study we regard $H(E)$ as ``flat" when all possible $H(E)$ is larger than $80\%$ of the average of $H(E)$.
Finally, $f$ becomes sufficiently small (e.g., $f \simeq 10^{-8}$), the simulation is stopped and the DOS is obtained.

In order to obtain temperature-composition phase diagrams on alloys, both the free energy and compositions are required at a specific temperature.
Under specific compositions, the Helmholtz free energy is obtained by the partition function; $F = -k_{\rm B} \ln Z$ at any temperature.
Finally, the phase diagrams is obtained by the Helmholtz free energy landscape.

The problem using canonical ensemble is that configuration attains to be a phase-separated mixture.
This causes such as interfacial contributions to total energy which lead to errors for calculating phase diagram.
This problem is avoided using semi-grand-canonical ensemble.

\subsection{Present extensiton of Wang-Landau algorithm for semi-grand-canonical ensemble}
\label{sec:wl_sgc}
Semi-grand-canonical (SGC) ensemble has been widely used for estimating phase diagram of alloys.
In SGC ensemble, concentrations are allowed to vary under externally imposed chemical potentials with the fixed total number of atoms.
This is different from grand-canonical ensemble where both concentrations and the total number of atoms vary.
One of the advantages of SGC is that configuration never attains to a phase-separated mixture\cite{0965-0393-10-5-304}.
This means that the configuration always shows a pure phase.
Therefore, interfacial contributions from a phase-separated mixture does not contribute to the calculated thermodynamic properties in SGC ensemble.

Hereinafter, for simplicity, we consider A-B binary alloy without lack of generality.
Therefore, we simply regard the concentration $x$ as $x = x_{\rm B}$ and the chemical potential $\mu$ as $\mu = \mu_{\rm B} - \mu_{\rm A}$.

In SGC ensemble, corresponding partition function is defined as:
\begin{equation}
	\label{eq:sgc}
	Y (T, \mu) = \sum_{E,x } W (E, x) \exp \left( -\frac{E-\mu x}{k_{\rm B}T} \right) .
\end{equation}
Unlike in canonical ensemble where the compositions are trivial to obtain, in SGC ensemble, since we handle chemical potentials instead of compositions, we should calculate ensemble averaged compositions.
A straightforward solution to obtain the compositions in SGC ensemble is
\begin{equation}
	\label{eq:sgc_straightforward}
	\langle x \rangle = \frac{ \sum_{E,x} x W(E,x) \exp \left( -\frac{ E - \mu x }{k_{\rm B}T} \right) } { Y (T, \mu) } .
\end{equation}
Here $\langle \rangle$ denotes the ensemble average.
Eq.~(\ref{eq:sgc_straightforward}) means that calculating phase diagrams on multicomponent alloy requires a multi-dimensional DOS in terms of total energy and compositions.
There are a lot of studies on calculating a multi-dimensional DOS\cite{Zhou_Wang_2006, Tsai_Uncovering_2008, Vogel_Generic_2013, Li_A_2014, Vogel_Exploring_2014, Vogel_Scalable_2014} using the WL algorithm.
The difficulties for calculating a multi-dimensional DOS are too time-consuming to construct the DOS and the error occurred in matching the piece of DOS that causes significant errors in thermodynamic properties.
Recently the successful parallel exchange scheme\cite{Vogel_Generic_2013, Li_A_2014, Vogel_Exploring_2014, Vogel_Scalable_2014} overcame these problems.
However, this parallel exchange method requires a lot of CPU cores and its implementation remains difficult.
In order to avoid these difficulties, we suggest another solution to use the thermodynamic relation.

If we consider $\mu$ as constant, total energy in Eq.~(\ref{eq:CE}) is rewritten as:
\begin{equation}
	\hat{E} = E-\mu x.
\end{equation}
Therefore, Eq.~(\ref{eq:sgc}) is rewritten as:
\begin{equation}
	\label{eq:sgc2}
	Y (T, \mu) = \sum_{\hat{E}} W_{\mu}(\hat{E}) \exp \left( -\frac{\hat{E}}{k_{\rm B}T} \right).
\end{equation}
The thermodynamic potential in SGC ensemble, $\phi$, is derived from $\phi(T, \mu) = -k_{\rm B} \ln Y$ likewise the Helmholtz free energy in canonical ensemble.
Otherwise, the ensemble averaged composition could not be calculated through Eq.~(\ref{eq:sgc2}) because of the lack of $x$ in the DOS.
Here, $\phi$ has a relation to $F$ through the Legendre transformation:
\begin{equation}
	\label{eq:legendre}
	\phi = F - \mu x.
\end{equation}
We can get $x$ using partial differentiation through the interpolating $\phi$ for each chemical potentials :
\begin{equation}
	\label{eq:composition_sgc}
	x = - \frac{\partial \phi}{\partial \mu}.
\end{equation}
Note that in SGC ensemble, since the concentrations are not fixed, the single-spin flip method can be used.
This also acceralates to sample rare states such as the ground state and is the advantage over the conventional WL algorithm in canonical ensemble.


An advantage of our method is not to need all the information of DOS.
If we would like to obtain free energies under a specific $\mu$ via the conventional WL algorithm, we should construct whole the multi-dimensional DOS.
However, constructing the multi-dimensional DOS is too difficult.
On the other hand, in our method, we need only a few one-dimensional DOS near the specific $\mu$.
This advantage enables us to obtain the temperature-composition dependent property under a specific $\mu$ in multicomponent alloys with high validity and accuracy which are lost in TI.

\section{\label{sec:level3} results and discussion}
In order to confirm validity and applicability of our method, we applied our method to Cu-Au (Sec.~\ref{sec:cuau}) and Pd-Rh (Sec.~\ref{sec:pdrh}) alloys.
Cu-Au alloy shows 1st order-disorder phase transition, and has been quite studied\cite{PhysRevB.57.6427, PhysRevB.44.512, PhysRevB.36.4163, PhysRevB.58.R5897} in terms of experiments and first-principles calculations.
Through the application to Cu-Au, we confirmed whether our method could describe the ordering tendency of alloys.
Pd-Rh alloy, on the other hand, shows the phase-separation tendency where two phases are coexist.
Since it is important to confirm whether our method could describe the phase coexistence and Pd-Rh is also quite a studied\cite{PhysRevB.37.5982, PhysRevLett.66.1753, PhysRevB.74.064202} system, we applied our method to the Pd-Rh alloy.

Note that in this study, for simplicity, we only consider configurational free energy, not include non-configurational free energy, e.g., vibrational and electronic free energy.
Although, in general, the vibrational effect is significant for phase diagrams especially in ordering alloys,
it just lowers the transition temperatures and does not change the low-temperature phases in Cu-Au\cite{PhysRevB.58.R5897}.
Therefore, even if we consider only the configurational free energy, we could capture the characteristics for ordering tendency and phase diagrams of Cu-Au  and phase-separation tendency and phase diagram of Pd-Rh without lack of validity.

\subsection{\label{sec:cuau} Cu-Au}

\begin{figure}
	\center
	\includegraphics[width=0.8\columnwidth]{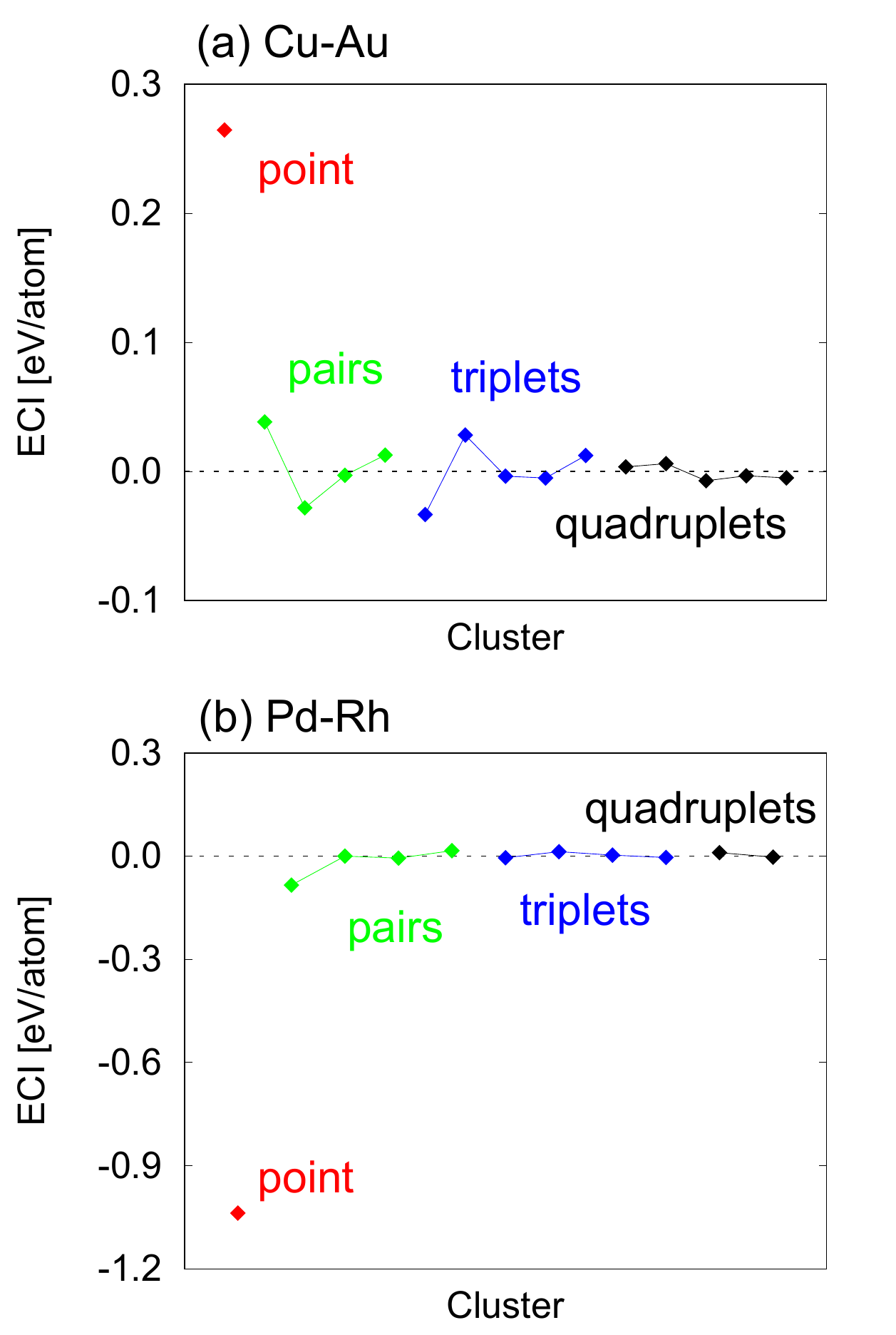}%
	\caption{\label{fig:eci}
		ECI for (a) Cu-Au and (b) Pd-Rh alloys except for empty cluster ECI, with the broken line indicating zero.
	}
\end{figure}

For calculation condition, total energies are obtained by the first-principles calculation via the VASP code \cite{PhysRevB.47.558, PhysRevB.54.11169}, based on the projector-augmented wave method (PAW) \cite{PhysRevB.59.1758} within the generalized-gradient approximation of Perdew-Burke-Ernzerhof (GGA-PBE) \cite{PhysRevLett.77.3865} to the exchange-correlation functional. The plane wave cutoff of 500 eV is used, and atomic positions are fully relaxed on underlying fcc lattice. Total energies of 183 structures consisting of up 32 atoms are calculated.
We obtained 16 optimized ECI (see Fig.~\ref{fig:eci}(a)) with the prediction accuracy, a cross-validation score, of 1.1 meV/atom, which gives sufficient accuracy to capture the thermodynamic characteristics for Cu-Au alloy.
In Fig.~\ref{fig:eci}(a), since nearest neighbor pair ECI has largest positive value, we can see the strong tendency to order.
The triplet and quadruplet ECIs, which mainly mean a contribution of atomic local relaxation, are large because there are large difference in lattice constant between Cu and Au.
\begin{figure}
	\center
	\includegraphics[width=0.8\columnwidth]{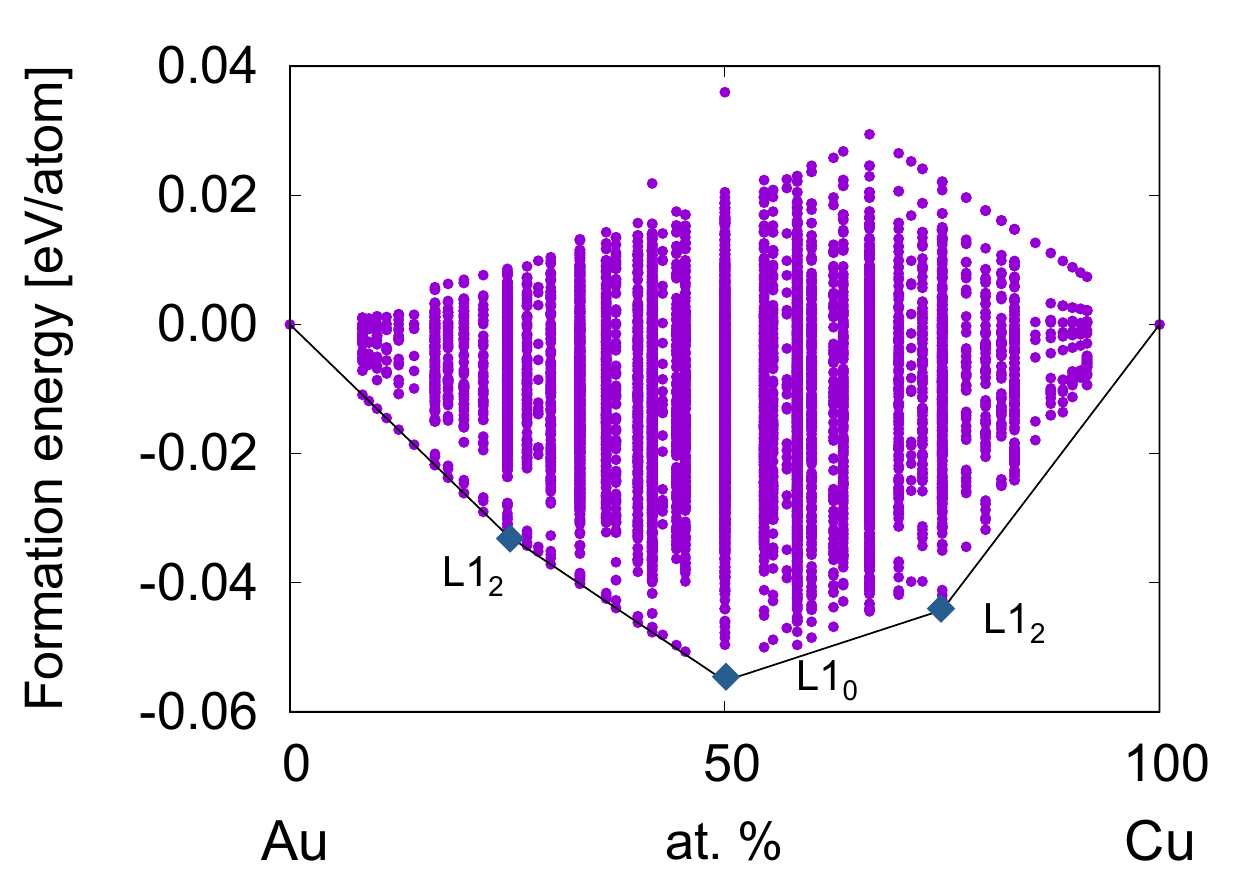}%
	\caption{\label{fig:formation_energies}
		Formation energies obtained via all derivative structures up to 12 atoms, 10850 structures.
	}
\end{figure}
In Fig.~\ref{fig:formation_energies}, through using derivative structures\cite{PhysRevB.77.224115} up to 12 atoms, we checked whether our ECIs are valid for describing $L1_0$ at ${\rm Cu}_{0.5}{\rm Au}_{0.5}$ and $L1_2$ at ${\rm Cu}_{0.75}{\rm Au}_{0.25}$ and ${\rm Cu}_{0.25}{\rm Au}_{0.75}$, and confirmed that our ECI completely describe low temperature phases in Cu-Au.

The simulations have been done for $4 \times 4 \times 4$ supercell on fcc until the factor attains $ f_{\rm final} \leq \exp(10^{-7}) $.
For each simulation, we set the chemical potential $ \mu = \mu_{\rm Au} - \mu_{\rm Cu} = $ 0.313-0.703 eV/atom, and calculate $\phi(T,\mu)$ using $W_\mu(\hat{E})$ and Eq.~(\ref{eq:sgc2}).
\begin{figure}
	\center
	\includegraphics[width=0.8\columnwidth]{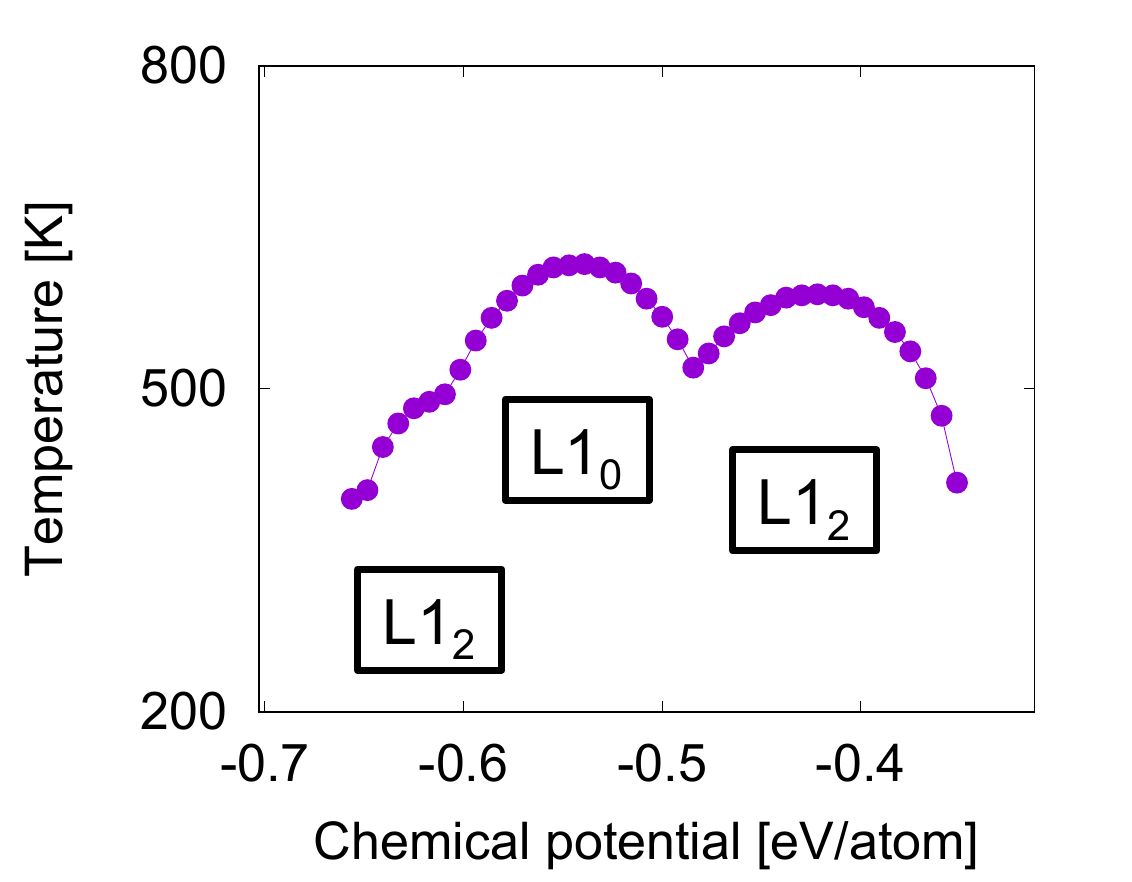}%
	\caption{\label{fig:bm_cuau}
		Cu-Au $T-\mu$ phase diagram. Points denote highest heat capacities for each chemical potential.
	}
\end{figure}
In Fig.~\ref{fig:bm_cuau}, we plot $T-\mu$ phase diagram where a phase transition point is regarded as one which shows the highest heat capacity, $C_{\rm max}(T, \mu)$.
We confirmed three order phases in Cu-Au, $L1_0$ for ${\rm Cu}_{0.5}{\rm Au}_{0.5}$ and $L1_2$ for both ${\rm Cu}_{0.75}{\rm Au}_{0.25}$ and ${\rm Cu}_{0.25}{\rm Au}_{0.75}$ through the whole of simulation.

\begin{figure}
	\center
	\includegraphics[width=1.0\columnwidth]{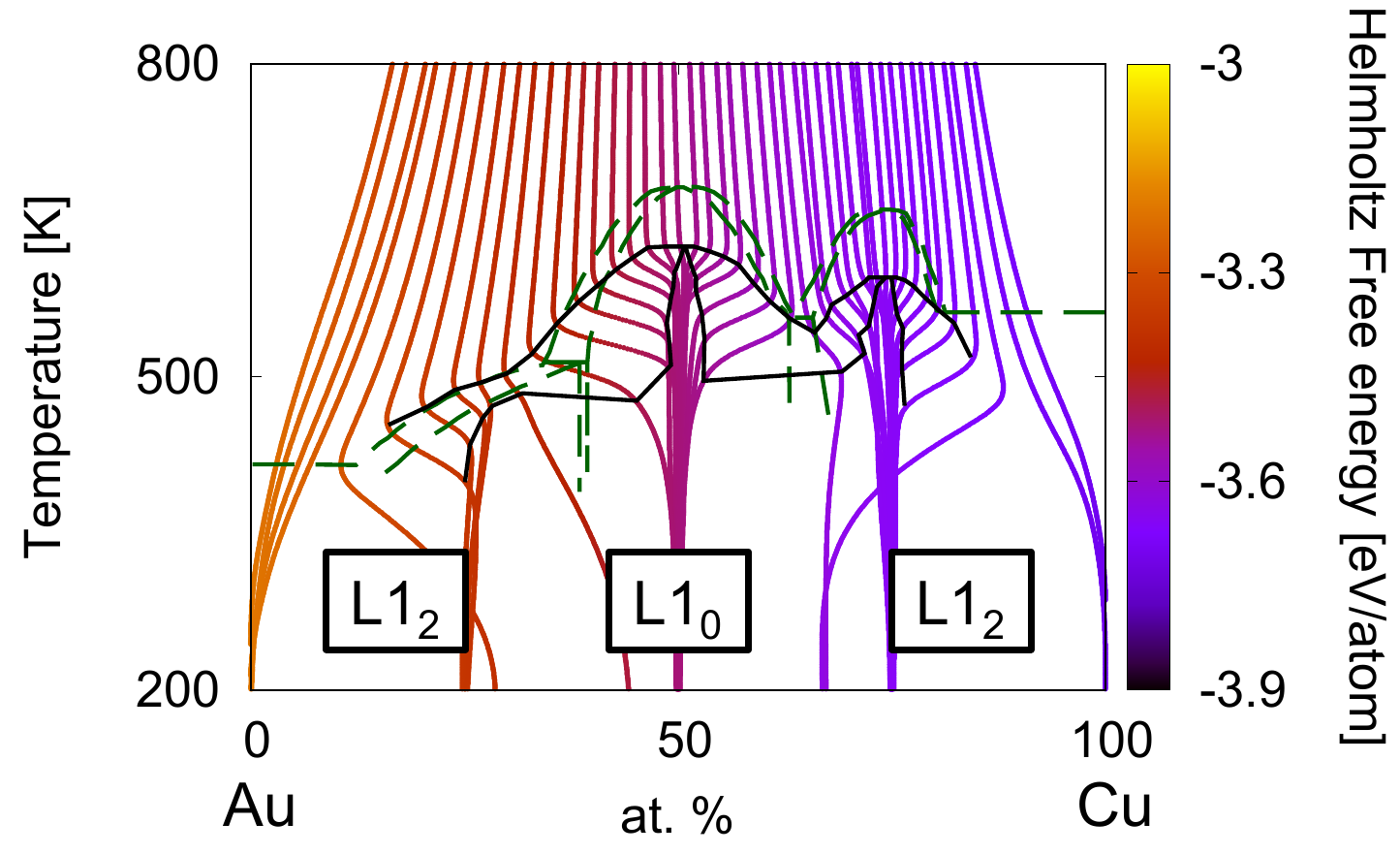}%
	\caption{\label{fig:phasediagram_cuau}
	The values of Helmholtz free energies and compositions obtained by our method over the range from 200K to 800K.
	Solid line denotes the order-disorder phase transition temperatures obtained by $T-\mu$ phase diagram in Fig.~\ref{fig:bm_cuau} and Eq.~(\ref{eq:composition_sgc}).
	Broken line denotes the result of experiment\cite{cuau-massalski}.
	}
\end{figure}

The temperature-composition phase diagram for Cu-Au by our method is shown in Fig.~\ref{fig:phasediagram_cuau}.
This phase diagram is obtained by converting $T-\mu$ phase diagram (Fig.~\ref{fig:bm_cuau}) into $T-x$ one (Fig.~\ref{fig:phasediagram_cuau}) through Eq.~(\ref{eq:composition_sgc}) with the interpolation of $\phi(T, \mu)$ for each chemical potential via cubic spline function.
Note that two phases are in equilibrium at a specific phase transition point in $T-\mu$ phase diagram.
Therefore, we consider one phase as one just before phase transition and another as just after that.
Compared with our result and the experimental one\cite{cuau-massalski}, we confirmed that our method successfully describe the Cu-Au phase diagram that captures the thermodynamic characteristics of Cu-Au alloy.
However, we could see that some simulation under a specific $\mu$ fails to describe $F(T,x)$ correctly.
This is because of the differential error in Eq.~(\ref{eq:composition_sgc}) and the smallness of our simulation cell.
As above, our method is found to be applicable to the alloy which shows ordering tendency.

\subsection{\label{sec:pdrh} Pd-Rh}
The calculation condition for Pd-Rh is almost same as that for Cu-Au in Sec.~\ref{sec:cuau}.
The different points are as follows; the plane wave cutoff of 600 eV is used and total energies of 71 structures consisting of up 32 atoms are calculated.
We obtained 12 optimized ECI (see Fig.~\ref{fig:eci}(b)) with a cross-validation score, of 0.8 meV/atom, which gives sufficient accuracy to describe the phase-separation tendency for Pd-Rh alloy.
In Fig.~\ref{fig:eci}(b), since nearest neighbor pair ECI has largest negative value, we can see the strong tendency to separate.
The triplet and quadruplet ECIs are small because there are little difference in lattice constant between Pd and Rh.

The simulations have been done for $6 \times 6 \times 6$ supercells on fcc until the factor attain $ f_{\rm final} \leq \exp(10^{-7}) $.
For each simulation, we set the chemical potential $\mu = \mu_{\rm Rh} - \mu_{\rm Pd} = $ (-2.079)-(-2.043) eV/atom.

\begin{figure}
	\center
	\includegraphics[width=1.0\columnwidth]{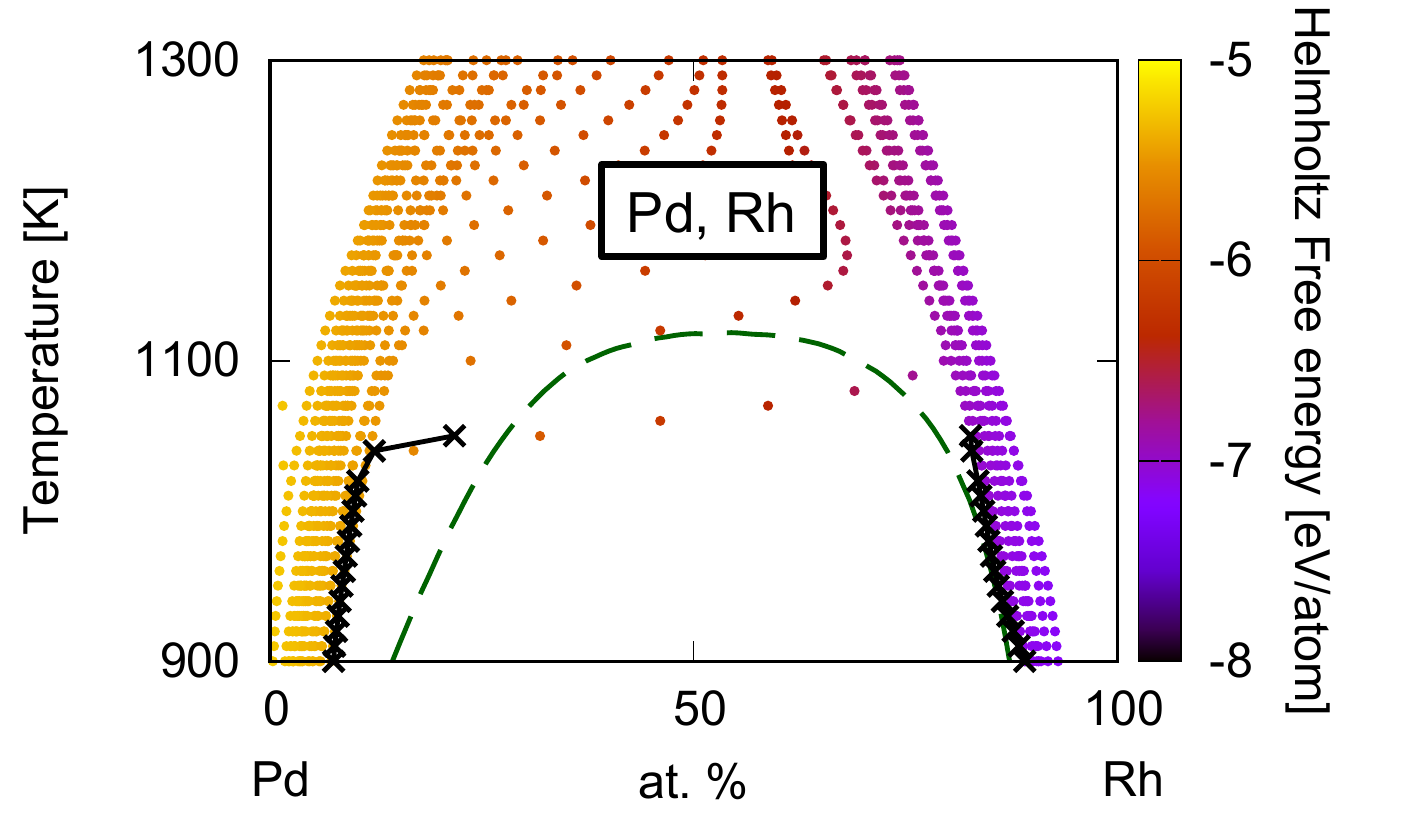}%
	\caption{\label{fig:phasediagram_pdrh}
	The values of Helmholtz free energies and compositions obtained by our method over the range from 900K to 1300K.
	Solid line denotes the phase-coexistent line obtained by our method.
	Broken line denotes the result of experiment\cite{pdrh-massalski}.
	}
\end{figure}
The temperature-composition phase diagram for Pd-Rh is shown in Fig.~\ref{fig:phasediagram_pdrh} with the same procedure as Cu-Au.
Unlike Cu-Au which shows 1st-order phase transition, the phase-separation system does not always show the clear peak of heat capacity under a specific $\mu$.
However, through comparing the Helmholtz free energy for each phase at a given temperature, we could detect the phase boundary.
\begin{figure}
	\center
	\includegraphics[width=0.8\columnwidth]{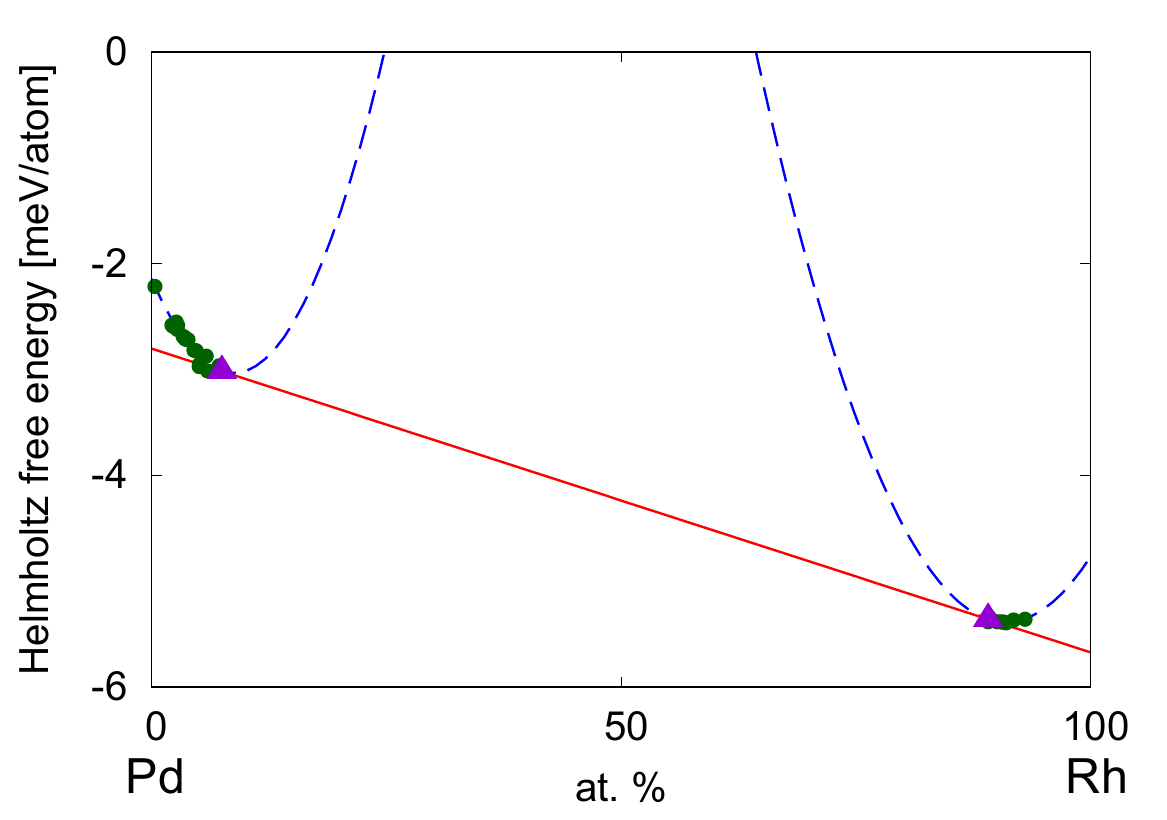}%
	\caption{\label{fig:helmholtz_f_T900}
	Closed circle points are the values of Helmholtz free energy at 900K.
	Quadratic curves denote the fitted value of points for each phase.
	A straight line denotes a common tangent line between the two quadratic curves.
	Closed triangle points denote the points of contact between the curves and the line.
	}
\end{figure}
In Fig.~\ref{fig:helmholtz_f_T900}, the Helmholtz free energy at $T=900$K is shown with a linear transformation for the sake of clarity.
With the least-square fitting to a quadratic curve for each phase, we clearly see a common tangent line between two curves that means the phase coexistence at a specific $T$ and $\mu$.
Regarding the phase boundary as the points of contact between the quadratic curves and the cotangent line,
we could explicitly detect the phase boundary for Pd-Rh that is shown in the solid line of Fig.~\ref{fig:phasediagram_pdrh}.

\section{\label{sec:level4} summary}
We suggest a new approach to obtain temperature-composition phase diagrams for multicomponent alloys using the Wand-Landau algorithm.
Since the advantage of our method does not suffer from the 1st- and 2nd-order phase transition,
we can calculate the phases and free energies with high accuracy even though near the phase transition points
where Metropolis algorithm and thermodynamic integration lose accuracy.
Through the application of our method to Cu-Au and Pd-Rh, we successfully obtained the phase diagrams and free energies.
This new approach shows that the we can replace the multi-dimensional DOS as multiple one-dimensional DOSs for estimating alloy phase diagram.

\begin{acknowledgments}
This work was supported by a Grant-in-Aid for Scientific Research (16K06704) from the MEXT of Japan, Research Grant from Hitachi Metals $\cdot$ Materials Science Foundation, and Advanced Low Carbon Technology Research and Development Program of the Japan Science and Technology Agency (JST).
\end{acknowledgments}

\bibliography{wl_sgc.bib}

\end{document}